\documentclass[aps,prd,showpacs,nofootinbib]{revtex4}
\usepackage{amsmath}
\usepackage{graphicx}
\usepackage{dcolumn}
\usepackage{amssymb}
\usepackage{color}

\def\lsim{\hbox{ \raise.35ex\rlap{$<$}\lower.6ex\hbox{$\sim$}\ }}
\def\gsim{\hbox{ \raise.35ex\rlap{$>$}\lower.6ex\hbox{$\sim$}\ }}

\begin{document}

\title{Scattering of cosmic strings by black holes: loop formation}
\author{Florian Dubath} \email{dubath@kitp.ucsb.edu}
\affiliation{Kavli Institute for Theoretical Physics, University of
California, Santa Barbara, CA 93106-4030, U.S.A.}

\author{Mairi Sakellariadou}
\email{Mairi.Sakellariadou@kcl.ac.uk}
\affiliation{Department of Physics, King's College, University of
London, Strand, London WC2R 2LS, United Kingdom}

\author{Claude Michel Viallet} \email{viallet@lpthe.jussieu.fr}
\affiliation{Laboratoire de Physique Th\'eorique et des Hautes
Energies, UMR 7589 CNRS~-~Paris~VI~-~Paris~VII, 4 place Jussieu, Boite
126, F-75252 Paris Cedex 05 France}

\date{\today}

\begin{abstract}
We study the deformation of a long cosmic string by a nearby rotating
black hole. We examine whether the deformation of a cosmic string,
induced by the gravitational field of a Kerr black hole, may lead to
the formation of a loop of cosmic string.  The segment of the string
which enters the ergosphere of a rotating black hole gets deformed
and, if it is sufficiently twisted, it can self-intersect chopping off
a loop of cosmic string. We find that the formation of a loop, via
this mechanism, is a rare event. It will only arise in a small region
of the collision phase space, which depends on the string velocity,
the impact parameter and the black hole angular momentum. We conclude
that generically, the cosmic string is simply scattered or
captured by the rotating black hole.

\end{abstract}

\pacs{04.20.-q, 04.70.Bw, 11.27.+d}
\maketitle

\section{Introduction}

Cosmic strings are one-dimensional topological
defects~\cite{td1,td2,td3} which could have been formed at the end of
phase transitions associated with spontaneously broken symmetries in
the early universe, via the Kibble mechanism~\cite{kibble}. Even
though cosmic strings do not play a dominant r\^ole in structure
formation, these objects are expected to be generically formed in the
framework of supersymmetric grand unified theories~\cite{jrs}.  Cosmic
microwave background measurements allow a small, but non negligible,
contribution of strings~\cite{cmbstr}, which can be realized in
supersymmetric inflationary models~\cite{johnm}. Recently, cosmic
strings gained a lot of interest since it was pointed out that
fundamental strings, D1-branes and D$(p-2)$-branes wrapping around
$(p-3)$-cycles in the compactified dimensions, could also play a
cosmic string-like r\^ole in cosmology~\cite{dstr}.

Cosmic string interactions can lead to a network of string loops.
Intercommutations of two long strings in two points, or
self-intercommutations of a long string in one point, may chop off a
loop of cosmic string. In what follows, we explore another mechanism
which may lead to the formation of cosmic string loops.  More
precisely, we examine the possibility that the scattering of a long
cosmic string by a rotating black hole can cause a deformation of the
long string, which may lead to the formation of a loop of cosmic
string. As a segment of a long cosmic string enters the ergosphere of
a rotating black hole, with rotation axis not parallel to the axis of
the cosmic string, this segment may be deformed and, in some cases, it
may get twisted. If the long string self-intersects, as a result of
the deformation, a loop of cosmic string may be formed.

The formation of loops of cosmic strings is important for the
astrophysical implications of a string network. In particular, cosmic
strings can be detected through the gravitational wave background
produced by oscillating string loops~\cite{dv}~\footnote{We
note that long cosmic strings also emit gravitational waves, since
these are wiggly due to string intercommutations~\cite{s}.}.

The scattering of a long cosmic string by a rotating black hole has
been previously studied in the case of a black hole whose axis of
rotation is parallel to the axis of the long string, concluding that
the long string will be either scattered, or captured by the black
hole. Which of the two regimes will take place depends on the impact
parameter. The critical value of the impact parameter which separates
the two regimes is given in Ref.~\cite{snajdr2}. The effect of the
scattering of a long cosmic string by a rotating black hole was
studied in Ref.~\cite{snajdr1}, where it was shown that the string is
displaced in the direction perpendicular to its velocity by an amount
which depends on the impact parameter~\cite{snajdr1}. In this article
we generalize the study of Ref.~\cite{snajdr1} for different
configurations of a long cosmic string approaching a rotating black
hole.

We conclude that the formation of a loop from the scattering of a long
string by a rotating black hole is a rare event. It will only arise in
a small region of the collision phase space, which depends on the
string velocity, the impact parameter and the black hole angular
momentum. More frequently, the cosmic string is either simply
scattered or captured by the rotating black hole.

In what follows we assume that the string thickness is negligible as
compared to its length. Moreover, the string linear mass density is
very small.  Then, assuming that the string size is much larger than
the Schwarzschild radius of the black hole, while the mass of the
string is much smaller than that of the black hole, we can treat the
cosmic string as a test object, and we can neglect its gravitational
back reaction.

The string is initially sufficiently far from the black hole, so it
can be taken as a straight string. As it approaches the black hole, it
will move in a Kerr geometry, however since the string is taken to be
sufficiently long, the end regions of the infinitely long string will
not be influenced by the presence of the gravitational field of the
Kerr black hole.

Let us consider the axis of rotation of the Kerr black hole to be
along the direction ${\cal X}$, in a system of coordinates
$({\cal X, Y, Z})$. A cosmic string being initially far away from the
black hole is moving towards it.  One can investigate the following
setups:
\begin{itemize}
\item
The string is at first parallel to the axis of rotation of the Kerr
black hole, and it moves towards the black hole with a velocity
perpendicular to the axis of rotation. Thus, the cosmic string is
originally along the ${\cal X}$-axis and its velocity is along the
${\cal Z}$-axis. [We call this case {\bf (i)}.]
\item The  string is at first perpendicular to the axis of rotation of
the Kerr black hole and it moves towards the black hole with velocity
parallel to axis of rotation. Thus, the cosmic string is originally
along the ${\cal Z}$-axis and its velocity is along the ${\cal
X}$-axis. [We call this case {\bf (ii)}.]
\item
The string is at first perpendicular to the axis of rotation of the
Kerr black hole and it moves towards the black hole with velocity
which is also perpendicular to the axis of rotation. Thus, the cosmic
string is originally along the ${\cal Z}$-axis and it moves towards
the Kerr black hole with velocity along the ${\cal Y}$-axis. [We call
this case {\bf (iii)}.]
\end{itemize}
 Case {\bf (i)} has been studied in Ref.~\cite{snajdr1}; in what
follows we will focus on cases {\bf (ii)} and {\bf (iii)}. Case {\bf
(iii)} is the most promising configuration to lead to the formation of
a cosmic string loop.

We organize the rest of the paper as follows: In Section II we briefly
discuss the string equations of motion in a curved background. We give
the string equations of motion for a string in a general background
geometry and then we derive them for a straight string far away from a
black hole. We proceed with first the Newtonian and then the
Lense-Thirring scattering of a straight string in the linearized
approximation. In Section III we study the scattering of a long
straight string by a Kerr black hole. Our numerical approach is
presented in Section IV. We round up with our conclusions
in Section V.

We will use units where $G=c=1$.

\section{String equations of motion in a curved background}

\subsection{String equations of motion}

The world history of a cosmic string can be expressed by a
two-dimensional surface in the four-dimensional space-time, which is
called the string worldsheet:
\begin{equation}
x^\mu=x^\mu(\zeta^a)~~;~~a=0,1~,
\end{equation}
where the worldsheet coordinates $\zeta^0, \zeta^1$ are arbitrary
parameters chosen so that $\zeta^0$ is time-like $(\equiv \tau$) and
$\zeta^1$ is space-like ($\equiv \sigma$).

The string equations of motion, in the limit of a zero thickness
string, are derived from the Nambu-Goto effective action which, up to
an overall factor, corresponds to the surface area swept out by the
string in space-time:
\begin{equation}
S_0[x^\mu]=-\mu\int\sqrt{-\gamma}d^2\zeta~,
\label{nga}
\end{equation}
where $\gamma$ is the determinant of the two-dimensional worldsheet
metric $\gamma_{ab}$:
\begin{equation}
\gamma={\rm det}(\gamma_{ab})=\frac{1}{2}
\epsilon^{ac}\epsilon^{bd}\gamma_{ab}\gamma_{cd}~~ ,
~~\gamma_{ab}=g_{\mu\nu}x^\mu_{,a}x^\nu_{,b}~,
\end{equation}
with $g_{\mu\nu}$ the four-dimensional metric.

One can derive the same string equations of motion by using Polyakov's
form of the action~\cite{pol}
\begin{equation}
S[x^\mu, h_{ab}]=-\frac{\mu}{2}\int
\sqrt{-h}h^{ab}\gamma_{ab}d^2\zeta~,
\label{spol}
\end{equation}
where $h_{ab}$ is the internal metric with determinant $h$.

Varying Eq.~(\ref{spol}) with respect to $x^\mu(\zeta^a)$ we obtain
the string equations of motion
\begin{eqnarray}
\Box x^\mu+h^{ab}\Gamma_{\nu\sigma}^\mu x^\nu_{,a}x^\sigma_{,b}&=&0
~~\mbox{(dynamical)}\nonumber\\ \gamma_{ab}-\frac{1}{2}
h_{ab}h^{cd}\gamma_{cd}&=&0~~\mbox{(constraint)}~,
\label{eom}
\end{eqnarray}
where
\begin{equation}
\Box=\frac{1}{\sqrt{-h}}\partial_a(\sqrt{-h}h^{ab}\partial_b)~,
\end{equation}
and $\Gamma^\mu_{\nu\sigma}$ is the four-dimensional Christoffel
symbol.  

We choose the gauge in which $h_{ab}$ is conformal to the flat two
dimensional metric $\eta_{ab}={\rm diag}(-1,1)$. The dynamical and
constraint equations, Eqs.~(\ref{eom}), become
\begin{eqnarray}
\Box x^\mu+ \eta^{ab}\Gamma_{\nu\sigma}^\mu
x^\nu_{,a}x^\sigma_{,b}&=&0\nonumber\\
\gamma_{01}=g_{\mu\nu}x^\mu_{,0}x^\nu_{,1}&=&0\nonumber\\
\gamma_{00}+\gamma_{11}=g_{\mu\nu}x^\mu_{,0}x^\nu_{,0}+x^\mu_{,1}
x^\nu_{,1}&=&0~,\label{eom_2}
\end{eqnarray}
where $\Box\equiv -\partial^2_\tau+\partial^2_\sigma$ (note that
$\partial_A\equiv \partial/\partial A$).

\subsection{Straight string moving in a weak gravitational field}

As long as a cosmic string is far away from a black hole, it just
feels a weak gravitational field and thus the linearized analysis is
sufficient. We present below the linear approach, which can be
employed to furnish the initial conditions allowing us to perform a
full numerical simulation for the case of a string being close to a
rotating black hole, and thus limiting the size of the numerical
simulation. This approach has been already used in Ref.~\cite{snajdr1}.

Assuming at first that there is no external gravitational field, the
4-dimensional space-time metric $g_{\mu\nu}$ is just the Minkowski
metric $\eta_{\mu\nu}$.  In  coordinates $({\cal T, X, Y,
Z})$ and choosing signature $(-,+,+,+)$ one can show that a solution
of the equations of motion Eq.(\ref{eom}) read
\begin{eqnarray}
x^\mu&=&x_0^\mu(\tau,\sigma)=(\tau\cosh(\beta),\tau\sinh(\beta)+{\cal
X}_0,b, \sigma)\nonumber\\ \mbox {or}~~
x^\mu&=&x_0^\mu(\tau,\sigma)=(\tau\cosh(\beta),b,\tau\sinh(\beta)+
{\cal Y}_0, \sigma)~,\nonumber\\ h_{ab}&=&\eta_{ab}={\rm diag}(-1,1)~.
\label{sol-eom1}
\end{eqnarray}
 Equation (\ref{sol-eom1}a) refers to a long straight string along
the ${\cal Z}$-axis, being initially at $x^\mu(0,\sigma)=(0,{\cal
X}_0,b,\sigma)$, and moving with velocity $v=\tanh(\beta)$ along the
direction ${\cal X}$. We consider $b>0, {\cal X}_0<0$. This
configuration corresponds to case {\bf (ii)}. Similarly,
Eq.~(\ref{sol-eom1}b) refers to a long straight string along the
${\cal Z}$-axis, being initially at $x^\mu(0,\sigma)=(0,b,{\cal
Y}_0,\sigma)$, and moving with velocity $v=\tanh(\beta)$ along the
direction ${\cal Y}$. We again consider $b>0, {\cal Y}_0<0$. This
configuration corresponds to case {\bf (iii)}.

As a second step we consider the case of a weak gravitational
field
\begin{equation}
g_{\mu\nu}=\eta_{\mu\nu}+\theta_{\mu\nu}~,
\end{equation}
in which moves a straight long string
\begin{equation}
x^\mu(\xi)=x_0^\mu(\xi)+{\bar x}^\mu(\xi)~;
\end{equation}
$\theta_{\mu\nu}, {\bar x}^\mu(\xi) $ denote a small metric
perturbation, and a small perturbation along a straight cosmic
string, respectively.

In first order in $\theta_{\mu\nu}$ and ${\bar x}^\mu$, the equations
of motion, Eq.(\ref{eom}), read
\begin{eqnarray}
\Box x^\mu+\Gamma_{\alpha\beta}^\mu(x_0)x_{0,a}^\alpha
x_{0,b}^\beta\eta^{ab}&=&0~~\mbox{(dynamical})\nonumber\\
\eta_{\mu\nu}\frac{\partial x_0^\mu}{\partial\tau}\frac{\partial {\bar
x}^\nu}{\partial\sigma} +\eta_{\mu\nu}\frac{\partial {\bar
x}^\mu}{\partial\tau}\frac{\partial x_0^\nu}{\partial\sigma}
+\theta_{\mu\nu}\frac{\partial x_0^\mu}{\partial\tau}\frac{\partial
x_0^\nu}{\partial\sigma} &=&0~~\mbox{(constraint})\nonumber\\
2\eta_{\mu\nu}\left(\frac{\partial
x_0^\mu}{\partial\tau}\frac{\partial {\bar x}^\nu}{\partial\tau}
+\frac{\partial x_0^\mu}{\partial\sigma}\frac{\partial {\bar
x}^\nu}{\partial\sigma}\right)+\theta_{\mu\nu} \left(\frac{\partial
x_0^\mu}{\partial\tau}\frac{\partial x_0^\nu}{\partial\tau}
+\frac{\partial x_0^\mu}{\partial\sigma}\frac{\partial
x_0^\nu}{\partial\sigma}\right)&=&0~~\mbox{(constraint})~.
\end{eqnarray}
The above equations describe the motion of a long
straight cosmic string, being located far from a Kerr black hole.

In the weak field approximation, the gravitational field produced by a
black hole of mass $M$ and angular momentum $J$, rotating around the
${\cal X}$-axis, is
\begin{equation}\label{ds_1}
ds^2=-\left(1-\frac{2M}{{\cal R}}\right)d{\cal
T}^2+\left(1+\frac{2M}{{\cal R}}\right) (d{\cal X}^2+d{\cal
Y}^2+d{\cal Z}^2)-\frac{4J}{{\cal R}^3}({\cal Y}d{\cal Z}-{\cal
Z}d{\cal Y})d{\cal T},
\end{equation}
where ${\cal R}^2={\cal X}^2+{\cal Y}^2+{\cal Z}^2$.

One can therefore write the small perturbation $\theta_{\mu\nu}$
on the gravitational field, in terms of the Newtonian and the
Lense-Thirring~\cite{kp,lt} parts, as
\begin{eqnarray}
\theta_{\mu\nu}&=&\theta_{\mu\nu}^{\rm N} + \theta_{\mu\nu}^{\rm
LT}\nonumber\\ &=&2\varphi\delta_{\mu\nu}+\frac{4J}{{\cal
R}^3}\delta^0_{(\mu}\epsilon_{\nu )\alpha 0 1}{\cal X}^\alpha~,
\label{ltpart}
\end{eqnarray}
where $\varphi$ is defined as $\varphi= M/{\cal R}$ and
$\epsilon_{\alpha\beta\gamma\delta}$ stands for the antisymmetric
tensor. The second term of the {\sl r.h.s} of Eq.~(\ref{ltpart}b) is
due to the rotating black hole around the ${\cal X}$-axis.

\subsection{Newtonian scattering of a straight string in the 
linearized approximation}

Let us first consider the Newtonian scattering of a straight cosmic
string. We first examine case {\bf (ii)}. The dynamical equations read
\begin{eqnarray}
\left(-\frac{\partial^2 }{\partial\tau^2} + \frac{\partial^2
}{\partial\sigma^2}\right)
{\bar x}^0&=&-2\sinh(\beta)\cosh(\beta)\varphi_{,1}\nonumber\\
\left(-\frac{\partial^2 }{\partial\tau^2} + \frac{\partial^2
}{\partial\sigma^2}\right){\bar x}^1&=&0\nonumber\\
\left(-\frac{\partial^2 }{\partial\tau^2} + \frac{\partial^2
}{\partial\sigma^2}\right)
{\bar x}^2&=&-2\sinh^2(\beta)\varphi_{,2}\nonumber\\
\left(-\frac{\partial^2 }{\partial\tau^2} + \frac{\partial^2
}{\partial\sigma^2}\right){\bar x}^3&=&-2\cosh^2(\beta)\varphi_{,3}~,
\end{eqnarray}
and the constraint equations are
\begin{eqnarray}
{\bar x}^3_{,\tau}-\cosh(\beta){\bar x}^0_{,\sigma}+\sinh(\beta){\bar
x}^1_{,\sigma}&= &0\nonumber\\ {\bar x}^3_{,\sigma}-\cosh(\beta){\bar
x}^0_{,\tau}+\sinh(\beta){\bar x}^1_{,\tau}&=
&-2\varphi\cosh^2(\beta)~;
\end{eqnarray}
we use the notation ${\cal W}_{,A}\equiv \partial {\cal W}/\partial
A$.

The dynamical equations imply that for a given time the cosmic
string lies in the $({\cal Y,Z})$-plane. The string deflection
${\cal D}$ along the plane orthogonal to the velocity of the
string, given in Ref.~\cite{snajdr1}, is
\begin{equation}
{\cal D}=b-2\pi M\sinh(\beta)~.
\end{equation}
It implies that while at later times segments of the cosmic string
which are far away from the rotating black hole keep moving in the
$({\cal Y}=b,{\cal Z})$-plane, since they are unaffected from the
gravitational field of the black hole, the central part of the string
being by definition closer to the location of the black hole will be
moving in the $({\cal Y}={\cal D},{\cal Z})$-plane instead. 

We then study case {\bf (iii)}.
The dynamical equations read
\begin{eqnarray}
\left(-\frac{\partial^2 }{\partial\tau^2} + \frac{\partial^2
}{\partial\sigma^2}\right)
{\bar x}^0&=&-2\sinh(\beta)\cosh(\beta)\varphi_{,1}\nonumber\\
\left(-\frac{\partial^2 }{\partial\tau^2} + \frac{\partial^2
}{\partial\sigma^2}\right){\bar x}^1&=&
-2\sinh^2(\beta)\varphi_{,1}
\nonumber\\
\left(-\frac{\partial^2 }{\partial\tau^2} + \frac{\partial^2
}{\partial\sigma^2}\right)
{\bar x}^2&=&0\nonumber\\
\left(-\frac{\partial^2 }{\partial\tau^2} + \frac{\partial^2
}{\partial\sigma^2}\right){\bar x}^3&=&-2\cosh^2(\beta)\varphi_{,3}~,
\end{eqnarray}
and the constraint equations are
\begin{eqnarray}
{\bar x}^3_{,\tau}-\cosh(\beta){\bar x}^0_{,\sigma}+\sinh(\beta){\bar
x}^2_{,\sigma}&= &0\nonumber\\ {\bar x}^3_{,\sigma}-\cosh(\beta){\bar
x}^0_{,\tau}+\sinh(\beta){\bar x}^2_{,\tau}&=
&-2\varphi\cosh^2(\beta)~.
\end{eqnarray}

\subsection{Lense-Thirring scattering of a straight string in
  the linearized approximation}

Let us now proceed with the Lense-Thirring scattering of a long
straight string, moving in the gravitational field of a black hole of
mass $M$ and angular momentum $J$, in the linearized approximation. 

Starting with case {\bf (ii)}, the dynamical equations read
\begin{eqnarray}
\left(-\frac{\partial^2 }{\partial\tau^2} + \frac{\partial^2
}{\partial\sigma^2}\right){\bar x}^0&=&\frac{6 Jb\sigma}{{\cal
R}^5}\nonumber\\ \left(-\frac{\partial^2 }{\partial\tau^2} +
\frac{\partial^2 }{\partial\sigma^2}\right){\bar x}^1&=&0\nonumber\\
\left(-\frac{\partial^2 }{\partial\tau^2} + \frac{\partial^2
}{\partial\sigma^2}\right){\bar x}^2&=& -\frac{6 J}{{\cal
R}^5}(\tau\sinh(\beta)+x_0)\sigma\cosh(\beta)\sinh(\beta)\nonumber\\
\left(-\frac{\partial^2 }{\partial\tau^2} + \frac{\partial^2
}{\partial\sigma^2}\right){\bar x}^3&=& \frac{6 J}{{\cal
R}^5}(\tau\sinh(\beta)+x_0)b\cosh(\beta)\sinh(\beta)~,
\end{eqnarray}
and  the constraint equations are
\begin{eqnarray}
{\bar x}^3_{,\sigma}-\cosh(\beta){\bar x}^0_{,\tau}+\sinh(\beta){\bar
  x}^1_{,\tau}&=&0\nonumber\\ -\cosh(\beta){\bar
  x}^0_{,\sigma}+\sinh(\beta){\bar
  x}^1_{,\sigma}+{\bar x}^3_{,\tau}&=&\frac{2J}{{\cal R}^3}b\cosh(\beta)~,
\end{eqnarray}
where
\begin{equation}{\cal R}=\sqrt{(\tau\sinh(\beta)+{\cal X}_0)^2+b^2+\sigma^2}~.
\end{equation}
Choosing initial conditions
\begin{eqnarray}
{\bar x}^0|_{\tau=0}&=&0\nonumber\\ {\bar
x}^2|_{\tau=0}&=&\frac{\partial {\bar
x}^2}{\partial\tau}|_{\tau=0}=0\nonumber\\ {\bar
x}^3|_{\tau=0}&=&\frac{\partial {\bar
x}^3}{\partial\tau}|_{\tau=0}=0~,
\end{eqnarray}
the dynamical equations can  be solved analytically. The solution reads
\begin{eqnarray}
{\bar x}^0&=&2Jb\Big[\frac{\sinh^2(\beta)\tau+{\cal
X}_0\sinh(\beta)-\sigma}{{\cal A}(\tau+\sigma){\cal R}}
-\frac{\sinh^2(\beta)\tau+{\cal X}_0\sinh(\beta)+\sigma}{{\cal
A}(\tau-\sigma){\cal R}} \nonumber\\ &&\ \ \ \ \
+\frac{\tau+\sigma-{\cal X}_0\sinh(\beta)}{{\cal A}(\tau+\sigma) {\cal
S}(\tau+\sigma)} - \frac{\tau-\sigma-{\cal X}_0\sinh(\beta)}{{\cal
A}(\tau-\sigma) {\cal S}(\tau-\sigma)} \Big]\nonumber\\
{\bar x}^2&=&2Jb\cosh(\beta)\Big[\frac{\sinh(\beta)[\sinh(\beta)
\{b^2+\sigma^2+\sigma\tau\}+\sigma {\cal X}_0]}{{\cal
A}(\tau+\sigma){\cal R}} -\frac{\sinh(\beta)[\sinh(\beta)
\{b^2+\sigma^2-\sigma\tau\}-\sigma {\cal X}_0]}{{\cal
A}(\tau-\sigma){\cal R}}\nonumber\\ &&
-\frac{\sinh(\beta)[\sinh(\beta) b^2+\sinh(\beta)(\tau+\sigma)^2+{\cal
X}_0(\tau+\sigma)]} {{\cal A}(\tau+\sigma){\cal S}(\tau+\sigma)}
+\frac{\sinh(\beta)[\sinh(\beta) b^2+\sinh(\beta)(\tau-\sigma)^2+{\cal
X}_0(\tau-\sigma)]} {{\cal A}(\tau-\sigma){\cal
S}(\tau-\sigma)}\nonumber\\
{\bar x}^3&=&2Jb\cosh(\beta)\Big[\frac{\sinh^2(\beta)\tau+{\cal
X}_0\sinh(\beta)+\sigma}{{\cal A}(\tau-\sigma){\cal R}}
+\frac{\sinh^2(\beta)\tau+{\cal X}_0\sinh(\beta)-\sigma}{{\cal
A}(\tau+\sigma){\cal R}} +\frac{\tau-\sigma-{\cal X}_0\sinh(\beta)}
{{\cal A}(\tau-\sigma){\cal S}(\tau-\sigma)}\nonumber\\ &&\ \ \ \ \ \
\ \ \ \ \ \ \ \ \ \ +\frac{\tau+\sigma-{\cal X}_0\sinh(\beta)} {{\cal
A}(\tau+\sigma){\cal S}(\tau+\sigma)} -\frac{\tau+\sigma}{({\cal
X}_0^2+{\cal Y}_0^2){\cal S}(\tau+\sigma)} - \frac{\tau-\sigma}{({\cal
X}_0^2+{\cal Y}_0^2){\cal S}(\tau-\sigma)} \Big]~,
\end{eqnarray}
where
\begin{eqnarray}
{\cal R}&=&\sqrt{\sinh^2(\beta)\tau^2+2\sinh(\beta)\tau{\cal
X}_0+{\cal X}^2_0+{\cal Y}_0^2+\sigma^2}\nonumber\\
{\cal S}(u)&=&\sqrt{{\cal X}_0^2+{\cal Y}_0^2+u^2}\nonumber\\
{\cal A}(u)&=&{\cal Y}_0^2[1+\sinh(\beta)^2]+[{\cal X}_0+\sinh(\beta)u]^2~.
\end{eqnarray}

Let us proceed with case {\bf (iii)}.
The dynamical equations read
\begin{eqnarray}
\left(-\frac{\partial^2 }{\partial\tau^2} + \frac{\partial^2
}{\partial\sigma^2}\right){\bar x}^0&=&-\frac{6 J\sigma}{{\cal
R}^5}(\tau \sinh(\beta)+y_0)\cosh^2(\beta)\nonumber\\
\left(-\frac{\partial^2 }{\partial\tau^2} + \frac{\partial^2
}{\partial\sigma^2}\right){\bar x}^1&=&\frac{6 Jb\sigma}{{\cal
R}^5}\cosh(\beta)\sinh(\beta)\nonumber\\
\left(-\frac{\partial^2 }{\partial\tau^2} + \frac{\partial^2
}{\partial\sigma^2}\right){\bar x}^2&=&0\nonumber\\
\left(-\frac{\partial^2 }{\partial\tau^2} + \frac{\partial^2
}{\partial\sigma^2}\right){\bar x}^3&=& \frac{2 J}{{\cal
R}^5}({\cal R}^2-3b^2)~,
\label{case3de}
\end{eqnarray}

and  the constraint equations are
\begin{eqnarray}
{\bar x}^3_{,\sigma}-\cosh(\beta){\bar x}^0_{,\tau}+\sinh(\beta){\bar
  x}^2_{,\tau}&=& -\frac{2J}{{\cal R}^3}\sigma\cosh(\beta)\sinh(\beta)
  \nonumber\\ -\cosh(\beta){\bar x}^0_{,\sigma}+\sinh(\beta){\bar
  x}^2_{,\sigma}+{\bar x}^3_{,\tau}&=&\frac{2J}{{\cal
  R}^3}b\cosh(\beta)~,
\label{case3c}
\end{eqnarray}
where
\begin{equation}{\cal R}=\sqrt{b^2+(\tau\sinh(\beta)+{\cal Y}_0)^2+\sigma^2}~.
\end{equation}

The solution is far more complicated than the one for case {\bf (ii)}\footnote{Instead
of writing down explicitly the quite tedious solution, we give in an appendix the
method we follow to solve the system of equations.}.

\section{Scattering of a straight string by a Kerr black hole}
We now turn to the full non-linear analysis. The Kerr metric, can be
presented in the {\sl quasi-Cartesian} coordinate system $({\cal
T,X,Y,Z})$ by using the following shift~\cite{snajdr1} from the
Boyer--Lindquist system $(t,r,\theta,\phi)$:
\begin{equation}
r={\cal R}+M~,
\end{equation}
and
\begin{eqnarray}
{\cal T}&=&t\nonumber\\
{\cal X}&=&{\cal R}\cos\theta\nonumber\\
{\cal Y}&=&{\cal R}\sin\theta\cos\phi\nonumber\\
{\cal Z}&=&{\cal R}\sin\theta\sin\phi~,
\label{qccs}
\end{eqnarray}
In these coordinates, the rotation axis of the black hole is the
${\cal X}$ axis and the metric takes the form
\begin{eqnarray}
ds^2&=& -\left(1-\frac{2M({\cal R}+M)}{{A^2}}\right)d{\cal T}^2
+\frac{4aM({\cal R}+M)}{A^2{\cal R}^2}\left({\cal Y}d{\cal Z}-{\cal
  Z}d{\cal Y}\right)d{\cal T}\nonumber\\ && 
  +\left(\frac{\left(({\cal
    R}+M)^2+a^2\right)^2-\frac{a^2\rho^2({\cal Y}^2+{\cal Z}^2)}{{\cal
      R}^2}}{({\cal Y}^2+{\cal Z}^2){\cal R}^2A^2}\right)\left({\cal
  Y}d{\cal Z}-{\cal Z}d{\cal Y}\right)^2 +\frac{A^2}{{\cal
    R}^2}\left(\frac{1}{\rho^2}+\frac{{\cal X}^2}{{\cal R}^2({\cal
    Y}^2+{\cal Z}^2)}\right)\left({\cal Y}d{\cal Y}+{\cal Z}d{\cal
  Z}\right)^2\nonumber\\ && 
  +\frac{A^2}{{\cal R}^2}\left(\frac{{\cal
    X}^2}{\rho^2}+\frac{{\cal Y}^2+{\cal Z}^2}{{\cal
    R}^2}\right)d{\cal X}^2 +\frac{2A^2}{{\cal
    R}^2}\left(\frac{1}{\rho^2}-\frac{1}{{\cal R}^2}\right)\left({\cal
  Y}d{\cal Y}+{\cal Z}d{\cal Z}\right){\cal X}d{\cal X},\label{ds_2}
\end{eqnarray}
where $M, J$ denote the mass and angular momentum, respectively,
of the black hole, withe $J=aM$ for $|a|\leq M$; $A$ and $\rho$
are defined as
\begin{equation}
A^2=\left({\cal R}+M\right)^2+\frac{a^2{\cal X}^2}{{\cal R}^2} \ \ \
{\rm and}\ \ \ \rho^2= \left({\cal R}+M\right)^2+a^2-2M\left({\cal
R}+M\right)\ .
\end{equation}
Far from the origin of the coordinate system, the metric takes the
form given in Eq.~(\ref{ds_1}).

\section{Numerical  approach}

We use the quasi-Cartesian coordinate system given previously.  To
study the case {\bf (ii)}, or {\bf (iii)}, we let evolve a string
initially parallel to ${\cal Z}$ axis with initial velocity along the
${\cal X}$ axis (case {\bf (ii)}), or the ${\cal Y}$ axis (case {\bf
(iii)}).  We focus on the more promising case of a maximally rotating
black hole ($a=M$) and we set the mass equal to unity; we can recover
other values by rescaling the coordinates.

The Kerr metric is given analytically in Eq.~(\ref{ds_2}) and the
Christoffel symbols are obtained by taking the derivatives of the
metric. In our simulations we compute the derivatives as a finite
difference for a small variation of the coordinates. We discretise
a piece of world-sheet with steps $\Delta\sigma$ and $\Delta\tau$.
The string position is obtained as the value of the fields
$X^\mu(\sigma,\tau)$ on the world-sheet. They are computed using
Eq.~(\ref{eom_2}) and performing a numerical integration along the
$\tau$ direction. We need initial and boundary conditions: the
initial conditions are the value and $\tau$ derivatives of
$X^\mu(\sigma,0)$, $V^\mu(\sigma,0)=\partial_\tau X^\mu(\sigma,0)$
on the slice $\tau=0$. We compute the boundary conditions as the
free evolution of the end of the world-sheet\footnote{ In a more accurate
simulation both initial and boundary conditions should be computed
using the weak field evolution.}.

In our simulations  the initial conditions are the values of the
$X^\mu(\sigma,0),V^\mu(\sigma,0)$ at the point $\sigma_i$,
$i\in\{0;...;4000\}$ given by
\begin{eqnarray}
X^\mu(\sigma_i,0)&=&(0,-{\mathcal X}_0,b,(i-2000)\delta {\mathcal Z})\
,\nonumber\\ V^\mu(\sigma_i,0)&=&(\cosh(\beta),\sinh(\beta),0,0)\
,\end{eqnarray} for the case {\bf (ii)}, and
\begin{eqnarray}
X^\mu(\sigma_i,0)&=&(0,b,-{\mathcal Y}_0,(i-2000)\delta {\mathcal Z})\
,\nonumber\\ V^\mu(\sigma_i,0)&=&(\cosh(\beta),0,\sinh(\beta),0)\ ,
\end{eqnarray}
for the case  {\bf (iii)}. In the above, $b$ is the impact parameter, $\delta
{\mathcal Z}=0.1/M$, $\beta$ denotes the velocity parameter and ${\mathcal
X}_0\ ({\mathcal Y}_0)$ is the distance between the cosmic
string and the black hole at the beginning of the simulation. In
order to avoid numerical instability we set the {\sl proper time step}
$\Delta\tau=10^{-3}$.

For a cosmic string which is deformed by a Kerr black hole but with
insufficient twist, so that a loop is not formed when the string comes
close to the black hole, we stop the simulation when the string starts
drifting away from the black hole. The reason for this is that the
deformation of the central segments of the cosmic string will be
propagated towards the end parts of the string, leaving behind a
wiggly, but on the large-scale straight, cosmic string.  We also stop
the simulation if we encounter an infinity, denoting that a point
along the string crosses the horizon of the black hole.

To display the string position we produce equal-time slices. Such
slices are obtained by searching all the world-sheet points
$(\sigma,\tau)$ such that $X^0(\sigma,\tau)={\mathcal T}$ for
fixed ${\mathcal T}$ and plotting the $({\mathcal X,Y,Z})$.

\begin{figure}[htbp]
\begin{center}
\includegraphics[width=11cm, height=7cm]{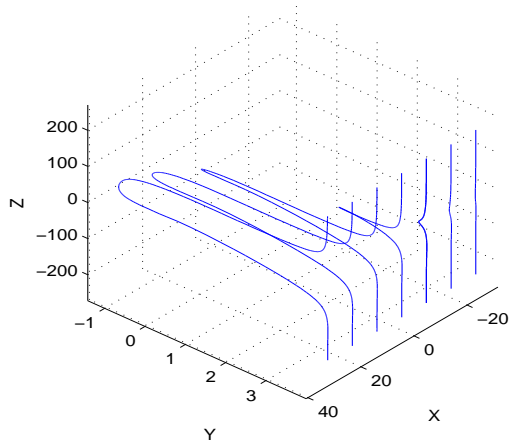}
\caption{Timeslices for the case {\bf (ii)} with $\beta=0.5$ and
$b=5$. We  show the slices for time ${\cal T}=25,37.5,\dots,100$. }
\label{II_evol}
\end{center}
\end{figure}

\begin{figure}[htbp]
\begin{center}
\includegraphics[width=7cm]{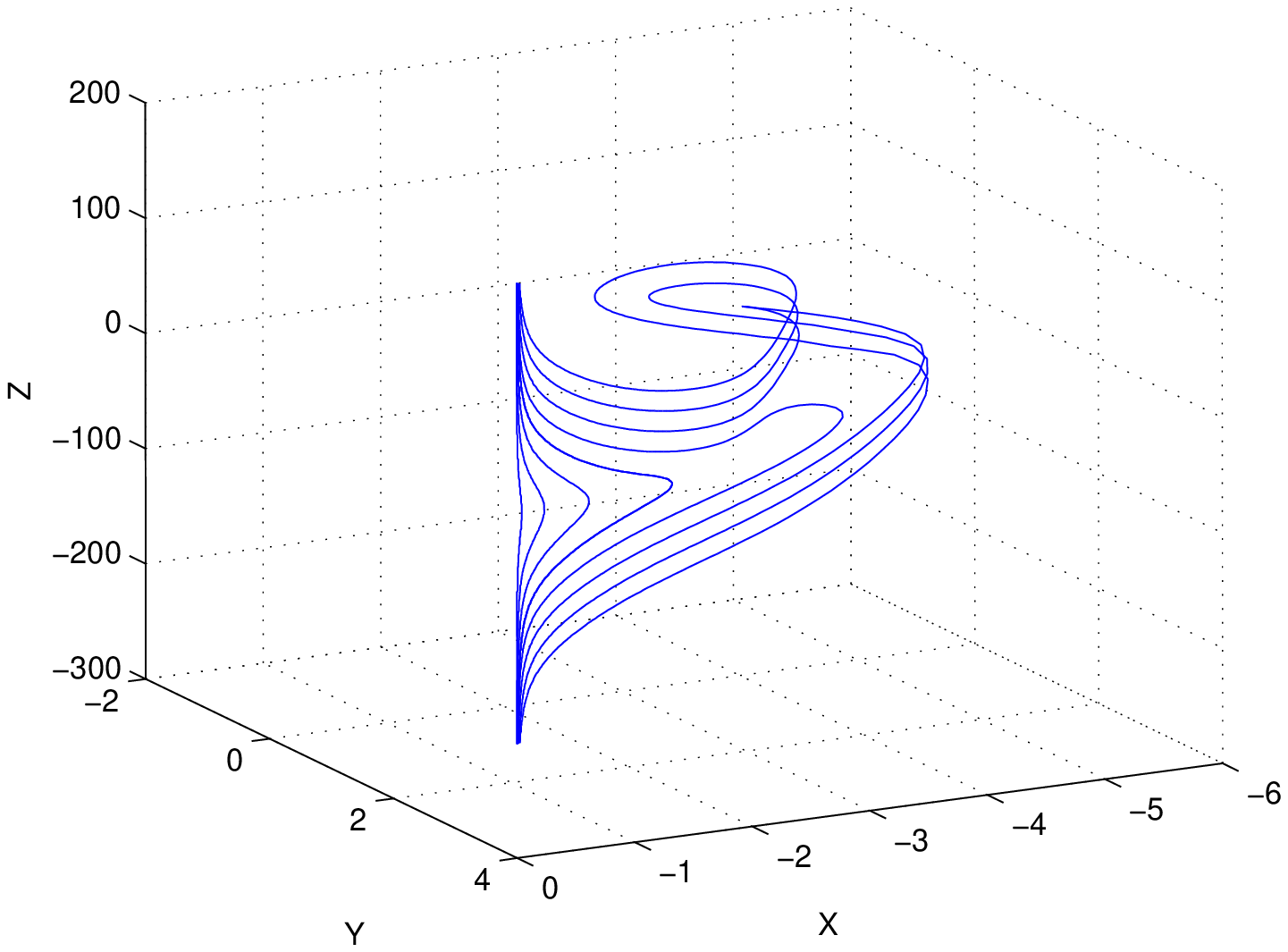}\includegraphics[width=7cm]{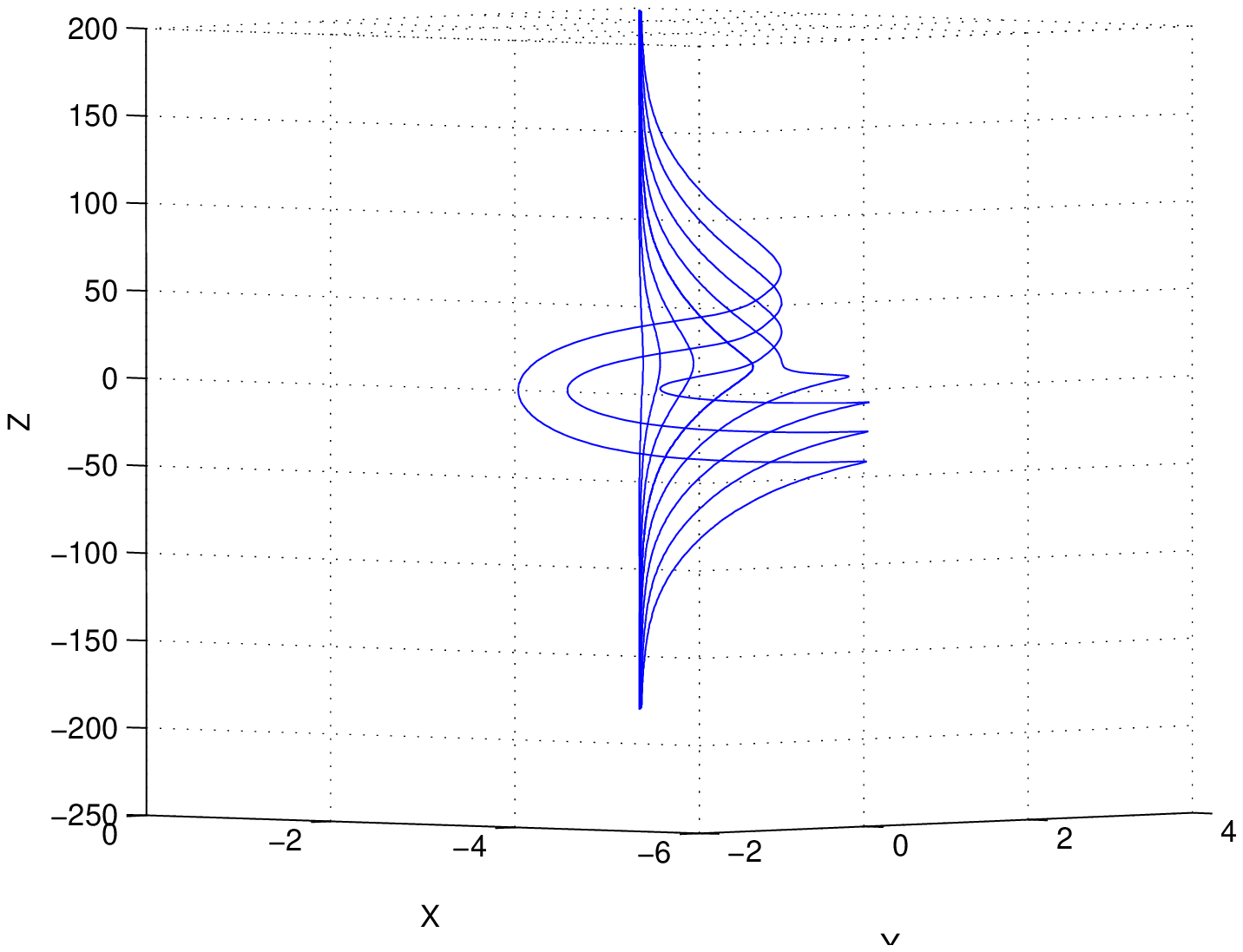}
\caption{Same as Fig.~\ref{II_evol} but with the ${\cal X}$ motion
subtracted in order to show the string deformation.   }
\label{II_def}
\end{center}
\end{figure}

\begin{figure}[htbp]
\begin{center}
\includegraphics[width=11cm, height=7cm]{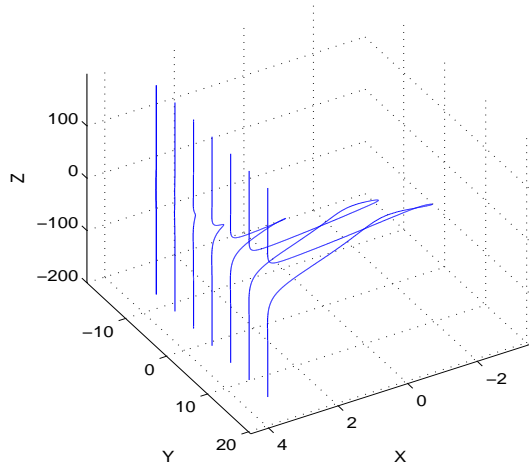}
\caption{Timeslices for the case {\bf (iii)} with $\beta=0.5$ and
$b=3.3$. We show the slices for time ${\cal
T}=44,55,\dots,110$. } \label{III_evol}
\end{center}
\end{figure}

\begin{figure}[htbp]
\begin{center}
\includegraphics[width=7cm]{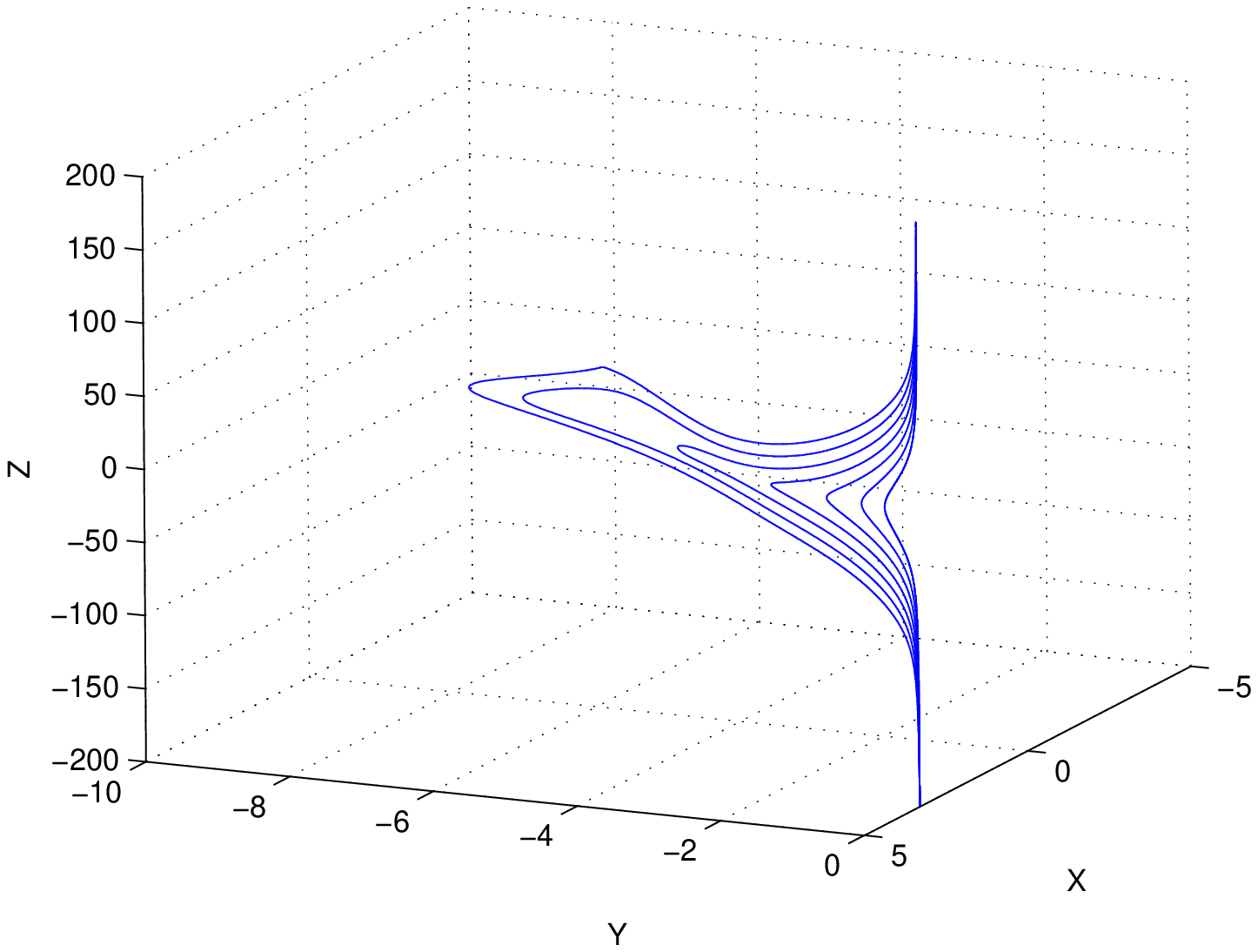}\includegraphics[width=7cm]{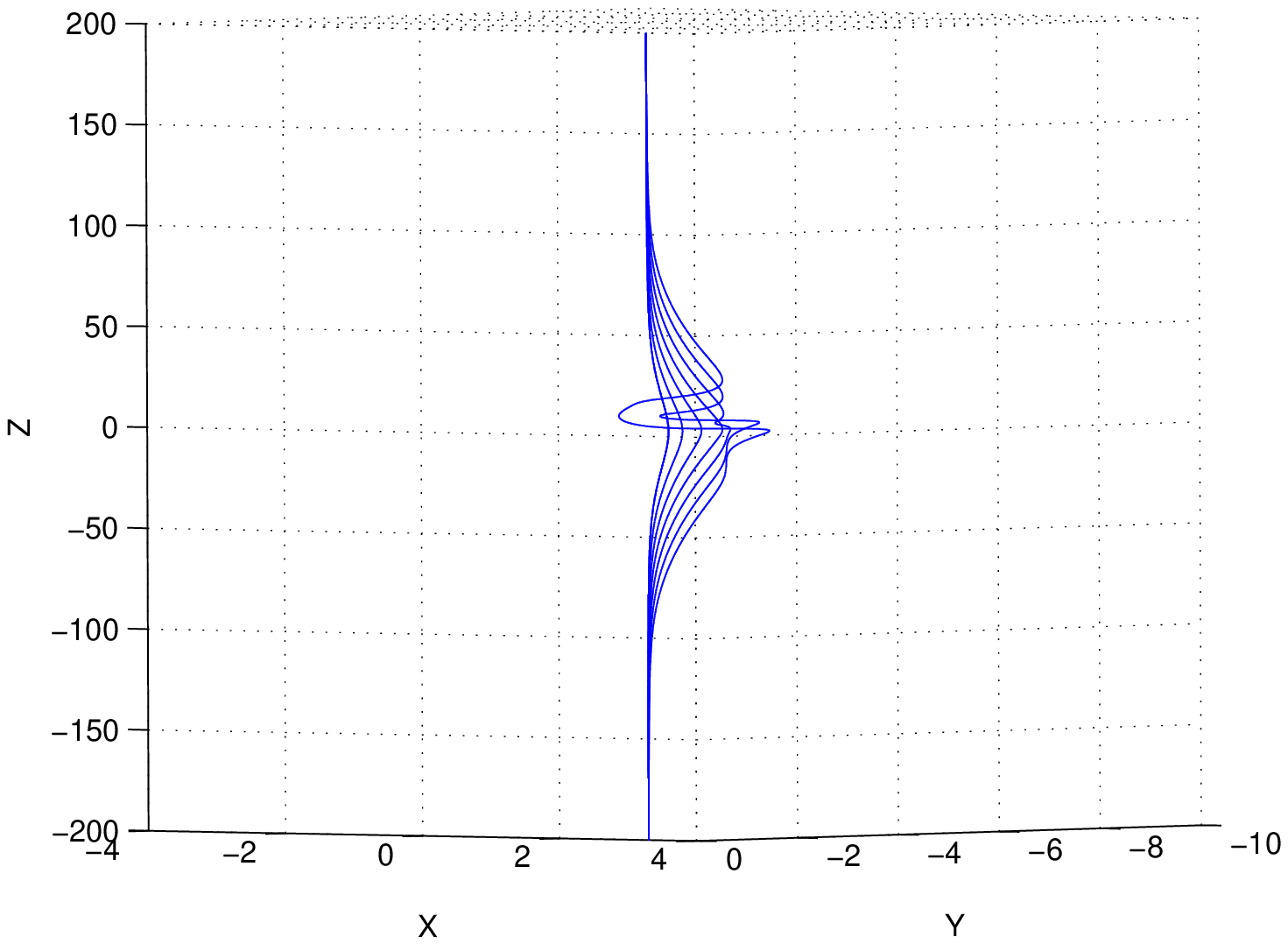}
\caption{Same as Fig. \ref{III_evol} but with the
${\cal Y}$ motion subtracted in order to show the string
deformation. Note that the center of the string has been twisted
of about 90 degrees.} \label{III_def}
\end{center}
\end{figure}

\section{Conclusions}

We have studied the deformation of a long cosmic string entering the
ergosphere of a Kerr black hole. The aim of this study is to
investigate how often such a deformation can be accompanied by a
sufficient twist, such that there is a string self-intersection
leading to a loop formation.

Among the configurations we have studied, the case of a long cosmic
string being at first perpendicular to the axis of rotation of the
Kerr black hole and moving towards it with velocity perpendicular to
the rotation axis, is the most prominent to lead to loop
formation. [It is our case {\bf (iii)}.]

A numerical investigation has shown us that loop formation is a rather
rare event, while in general the cosmic string will be just scattered
from the black hole or captured by it.

\acknowledgments It is a pleasure to thank Martin Snajdr for useful
discussions. The work of F.D. is supported by the Swiss National Funds
and by the National Science Foundation under Grant No. PHY99-07949.

\section*{Appendix}
Rather than giving explicitely the lengthy formulae for the solutions
of the dynamical equations, Eq.~(\ref{case3de}), of case {\bf (iii)}
with constraint equations Eq.~(\ref{case3c}), we describe here the
resolution method of such equations.

The form of the equations is 
\begin{eqnarray}
\label{eqnohom}
\partial_\sigma^2 \varphi(\sigma, \tau) - \partial_\tau^2
\varphi(\sigma, \tau) = f(\sigma, \tau)~.
\end{eqnarray}

The corresponding homogeneous equation is 
\begin{eqnarray}
\label{eqhom}
\partial_\sigma^2 \varphi_{\rm h}- \partial_\tau^2 \varphi_{\rm h}= 0~.
\end{eqnarray}

The general solution of Eq.~(\ref{eqnohom}) is the sum of the general
solution $\varphi_h$ of the homogeneous equation Eq.~(\ref{eqhom}) and a
particular solution $\varphi_{part}$ of Eq.~(\ref{eqnohom}).

The general solution $\varphi_h$ is just
\begin{eqnarray}
\label{solhom}
\varphi_{\rm h}( \sigma, \tau) = a\,(\sigma+\tau) + b\, ( \sigma- \tau)~,
\end{eqnarray}
wih arbitrary functions of  one variable $a$ and $b$.

It is possible to write a particular solution $\varphi_{\rm part}$ as a
double integral of the right hand side of (\ref{eqnohom}).
\begin{eqnarray}
\label{solpart}
\varphi_{\rm part} = \int_\Delta f(x,y)\; dx \; dy~,
\end{eqnarray}
where the domain of integration $\Delta$ is a triangle shown in the figure below.

\setlength{\unitlength}{2000sp}%
\begingroup\makeatletter\ifx\SetFigFont\undefined%
\gdef\SetFigFont#1#2#3#4#5{%
  \reset@font\fontsize{#1}{#2pt}%
  \fontfamily{#3}\fontseries{#4}\fontshape{#5}%
  \selectfont}%
\fi\endgroup
\bigskip
\begin{center}
\begin{picture}(5019,5919)(3139,-6373)
\thinlines {\color[rgb]{0,0,0}\put(3151,-4111){\vector( 1, 0){4995}}
}%
{\color[rgb]{0,0,0}\put(4051,-1861){\line( 1,-1){1800}}
\put(5851,-3661){\line(-1,-1){1800}} \put(4051,-5461){\line( 0,
1){3600}} \put(4051,-1861){\line( 0, 1){ 0}} }%
{\color[rgb]{0,0,0}\put(5851,-3661){\line( 0,-1){450}} }%
{\color[rgb]{0,0,0}\put(4051,-1456){\vector( 0, 1){990}}
\put(4051,-466){\line( 0,-1){5895}} }%
\put(5800,-4491){\makebox(0,0)[lb]{\smash{{\SetFigFont{16}{24.0}{\rmdefault}{\mddefault}{\updefault}{$\tau$}%
}}}}
\put(2700,-1771){\makebox(0,0)[lb]{\smash{{\SetFigFont{16}{24.0}{\rmdefault}{\mddefault}{\updefault}{$\sigma+\tau$}%
}}}}
\put(2700,-5461){\makebox(0,0)[lb]{\smash{{\SetFigFont{16}{24.0}{\rmdefault}{\mddefault}{\updefault}{$\sigma-\tau$}%
}}}}
\put(8200,-4281){\makebox(0,0)[lb]{\smash{{\SetFigFont{20}{24.0}{\rmdefault}{\mddefault}{\updefault}{x}%
}}}}
\put(4200,-871){\makebox(0,0)[lb]{\smash{{\SetFigFont{20}{24.0}{\rmdefault}{\mddefault}{\updefault}{y}%
}}}}
\put(5941,-3571){\makebox(0,0)[lb]{\smash{{\SetFigFont{20}{24.0}{\rmdefault}{\mddefault}{\updefault}{($\sigma$,$\tau$)}%
}}}}
\put(4501,-3481){\makebox(0,0)[lb]{\smash{{\SetFigFont{20}{24.0}{\rmdefault}{\mddefault}{\updefault}{$\Delta$}%
}}}}
\end{picture}
\end{center}

The particular solution $\varphi_{\rm part}$ may be written in more than one way.
Two are useful:

\begin{eqnarray}
\label{way1}
\varphi_{\rm part}(\sigma,\tau) = {1\over{2}}
\int_{\sigma-\tau}^\sigma dy\; \int_0^{\tau-\sigma+y} dx \; f(x,y) +
{1\over{2}} \int_\sigma^{\sigma+\tau} dy \; \int_0^{\tau+\sigma-y}
f(x,y)
\end{eqnarray}

and

\begin{eqnarray}
\label{way2}
\varphi_{\rm part}(\sigma,\tau) = {1\over{2}} \int_0^\tau dx \;
\int_{\sigma- \tau+x}^{\sigma+\tau-x} dy\; f(x,y)
\end{eqnarray}

Expression (\ref{way1}) [resp. (\ref{way2})] is useful when $f(x,y)$
may be written simply as a derivative $\partial_x \, g(x,y)$, [resp.
$\partial_y \, g(x,y)$]. Only one integration is then to be performed.

Of course equation (\ref{eqnohom}) is complemented with boundary
conditions which will determine the unknown functions $a$ and $b$.

\end{document}